\documentclass[a4paper,11pt]{article}
\usepackage{pos}

\title{Emergence of the rho resonance from the HAL QCD potential}
\author*[a,b]{Yutaro Akahoshi}
\author[a,b]{Sinya Aoki}
\author[b,c]{Takumi Doi}

\affiliation[a]{Center for Gravitational Physics, Yukawa Institute for Theoretical Physics, Kyoto University, \\
Kyoto 606-8502, Japan}

\affiliation[b]{RIKEN Nishina Center(RNC), \\
Saitama 351-0198, Japan}
\affiliation[c]{RIKEN Interdisciplinary Theoretical and Mathematical Sciences Program (iTHEMS), \\
Saitama 351-0198, Japan}

\emailAdd{yutaro.akahoshi@yukawa.kyoto-u.ac.jp}
\emailAdd{saoki@yukawa.kyoto-u.ac.jp}
\emailAdd{doi@ribf.riken.jp}

\abstract{
In this article, we report the $\rho$ resonance study using the HAL QCD method. We calculate the $I=1$ $\pi \pi$ potential at $m_{\pi} \approx 0.41$ GeV by a combination of the one-end trick, sequential propagator and covariant approximation averaging (CAA).
Thanks to those techniques, we determine the non-local $I=1$ $\pi \pi$ potential at the next-to-next-to-leading order (N$^2$LO) of the derivative expansion for the first time and obtain the pole of the S-matrix corresponding to the $\rho$ resonance.
We also discuss the comparison between our result and a previous calculation, done by L\"uscher's method.
}

\FullConference{%
 The 38th International Symposium on Lattice Field Theory, LATTICE2021
  26th-30th July, 2021
  Zoom/Gather@Massachusetts Institute of Technology
}

\begin{document}
\raisebox{80pt}[0pt][0pt]{\hspace*{50mm} YITP-21-146, RIKEN-QHP-511, RIKEN-iTHEMS-Report-21}
\vspace{-12mm}
\maketitle

\section{Introduction}
Understanding the hadronic resonances from the first-principle lattice QCD simulation is one of the most important subjects in particle and nuclear physics.
The HAL QCD method~\cite{Ishii:2006ec,Aoki:2009ji,Aoki:2011ep,HALQCD:2012aa}, which directly constructs inter-hadron potentials from spatial and temporal correlation functions calculated in lattice QCD,
is a way to shed light on the nature of those resonances with less systematic uncertainties
since we can directly study the pole structure of the S-matrix without any model-dependent ansatz.
There exists, however, a practical challenge for expensive computations of all-to-all quark propagators in resonance studies.
To overcome this difficulty, we have previously explored one of the all-to-all techniques, the hybrid method~\cite{Foley:2005ac},
but it is revealed that the numerical cost for the noise reductions is too large to employ simulations with larger lattice sizes~\cite{Akahoshi:2020ojo,Akahoshi:2019klc}.

In this study, we introduce a new strategy to handle all-to-all propagators where we combine three techniques: the one-end trick~\cite{McNeile:2006bz}, the sequential propagator calculation~\cite{Martinelli:1988rr}, and the covariant approximation averaging (CAA)~\cite{Shintani:2014vja}.
We calculate the HAL QCD potential of the $I=1$ $\pi \pi$ interaction on gauge configurations at $m_{\pi} \approx 0.41$ GeV, where the $\rho$ meson is known to appear as a resonance with $m_{\rho} \approx 0.89$ GeV.
The new strategy leads us to extract the potential at the next-to-next-to-leading order (N$^2$LO) in the derivative expansion for the first time.
We study a pole structure of the S-matrix using the N$^2$LO potential and find a pole corresponding to the $\rho$ resonance.
We also discuss a comparison between our result and the previous one obtained by L\"uscher's finite-volume method~\cite{Aoki:2011yj}.
Details of our analysis are covered in the published paper~\cite{Akahoshi:2021sxc}.

\section{HAL QCD method}
The fundamental quantity in the HAL QCD method is the Nambu--Bethe--Salpeter (NBS) wave function, which is defined as
\begin{equation}
  \psi_{W}({\bf r}) = \langle 0| (\pi \pi)_{I=1,I_z=0}({\bf r},0) |\pi \pi;I=1,I_z=0,{\bf k} \rangle,
\end{equation}
where $|\pi \pi;I=1,I_z=0,{\bf k} \rangle$ is an asymptotic state for an elastic $I=1$ $\pi \pi$ system in the center-of-mass frame with a relative momentum ${\bf k}$, a total energy $W = 2 \sqrt{m_{\pi}^2 + k^2}$ and $ k = \vert {\bf k} \vert$.
A sink operator $(\pi \pi)_{I=1,I_z=0}({\bf r},t)$ is a two-pion operator projected to the $I=1, I_z=0$ channel given by
\begin{eqnarray} \label{eq:sinkop}
  (\pi \pi)_{I=1,I_z=0}({\bf r},t) &=& \frac{1}{\sqrt{2}} \{ \pi_s^{+}({\bf r+x},t) \pi_s^{-}({\bf x},t) - \pi_s^{-}({\bf r+x},t) \pi_s^{+}({\bf x},t) \}, \\
  \pi_s^{+}({\bf x},t) &=& \bar d_s ({\bf x},t) \gamma_5 u_s ({\bf x},t), \quad
  \pi_s^{-}({\bf x},t) = \bar u_s ({\bf x},t) \gamma_5 d_s ({\bf x},t),
\end{eqnarray}
where $q_s ({\bf x},t) = \sum_{\bf y} f(|{\bf x - y}|) q({\bf y},t)$ for $q = \{u,d\}$ are slightly-smeared up and down quark fields with the smearing function $f(r) \equiv \{ e^{-r}, 1, 0 \}$ for $\{ 0 < r < 3.5, r = 0, 3.5 \leq r \}$ in lattice unit.
{
In general, physical observables are independent of a choice of the sink operator $(\pi \pi)_{I=1,I_z=0}({\bf r},t)$,
thus one can choose a convenient operator.
In our case, we employ the slightly smeared-sink operator (eq.(\ref{eq:sinkop})) since it reduces systematic uncertainty by making the potential in a short-range region smoother.
}

We extract the potential from the normalized correlation function $R({\bf r},t)$, which is a sum of NBS wave functions as
\begin{equation}
  R({\bf r},t) \equiv \frac{\langle 0| (\pi \pi)_{I=1,I_z=0}({\bf r},t) \overline{ {\mathcal J} }^{T_1^-}_{I=1,I_z=0}(t_0) |0 \rangle}{F_{\pi}(t-t_0)^2} \approx \sum_n A_n \psi_{W_n}({\bf r}) e^{-(W_n - 2m_{\pi}) (t-t_0)} + ...\ ,
\end{equation}
where $F_{\pi}(t)$ is the pion two-point function, $\overline{ {\mathcal J} }^{T_1^-}_{I=1,I_z=0}$ is a source operator which creates $I=1$ $\pi \pi$ states in $T_1^-$ representation, $W_n$ is the energy of the $n$th elastic state and an ellipsis indicates inelastic contributions.
For the source operators, we choose $\rho$-type $\overline{ {\mathcal J} }^{T_1^-}_{\rho, I=1,I_z=0}(t_0)$ and $\pi\pi$-type $\overline{ {\mathcal J} }^{T_1^-}_{\pi\pi, I=1,I_z=0}(t_0)$ in this study, defined by
\begin{eqnarray}
  \overline{ {\mathcal J} }^{T_1^-}_{\rho, I=1,I_z=0}(t_0) &=& \sum_{{\bf x}} \frac{1}{\sqrt{2}} \left(  \bar u({\bf x},t_0) \gamma_{3} u({\bf x},t_0) - \bar d({\bf x},t_0) \gamma_{3} d({\bf x},t_0) \right),\\
  \overline{ {\mathcal J} }^{T_1^-}_{\pi\pi, I=1,I_z=0}(t_0) &=& \frac{1}{\sqrt{2}} \sum_{{\bf y_1,y_2}} e^{-i{\bf p_3 \cdot y_1}} e^{i{\bf p_3 \cdot y_2}} \left( \pi^{-}({\bf y_1},t_0)\pi^{+}({\bf y_2},t_0) - \pi^{+}({\bf y_1},t_0)\pi^{-}({\bf y_2},t_0) \right),
\end{eqnarray}
\if0
\begin{eqnarray}
  \overline{ {\mathcal J} }^{T_1^-}_{\rho, I=1,I_z=0}(t_0) &=& \overline{\rho}^{0}_3 (t_0),\\
  \overline{ {\mathcal J} }^{T_1^-}_{\pi\pi, I=1,I_z=0}(t_0) &=&
  \overline{(\pi \pi)}_{I=1,I_z=0}({\bf p}_3,t_0) ,
\end{eqnarray}
\fi
where ${\bf p}_3 = (0,0,2\pi/L)$. We use local quark fields for source operators.
\if0
$\overline{(\pi\pi)}_{I=1,I_z=0}({\bf p},t)$ and $\overline{\rho}^{0}_3$ are given as
\begin{eqnarray}
  \overline{\rho}^{0}_3 (t) &=& \sum_{{\bf x}} \frac{1}{\sqrt{2}} \left(  \bar u({\bf x},t) \gamma_{3} u({\bf x},t) - \bar d({\bf x},t) \gamma_{3} d({\bf x},t) \right)\\
  \overline{(\pi\pi)}_{I=1,I_z=0}({\bf p},t) &=& \frac{1}{\sqrt{2}} \sum_{{\bf y_1,y_2}} e^{-i{\bf p \cdot y_1}} e^{i{\bf p \cdot y_2}} \left( \pi^{-}({\bf y_1},t)\pi^{+}({\bf y_2},t) - \pi^{+}({\bf y_1},t)\pi^{-}({\bf y_2},t) \right),
\end{eqnarray}
where we use local quark fields for source operators.
\fi

By using an asymptotic behavior of the NBS wave function\cite{Aoki:2009ji}, we can define an energy-independent non-local potential as\cite{HALQCD:2012aa}
\begin{equation}
  \left[ \frac{\nabla^2}{m_{\pi}} -\frac{\partial}{\partial t} + \frac{1}{4m_{\pi}} \frac{\partial^2}{\partial t^2} \right] R({\bf r},t) = \int d^3{\bf r'} U({\bf r},{\bf r'}) R({\bf r'},t),
\end{equation}
where the non-locality of the potential is treated by the derivative expansion $U({\bf r},{\bf r'}) = (V_0(r) + V_2(r) \nabla^2 + {\mathcal O}(\nabla^4)) \delta({\bf r-r'})$ in practice.
\if0
In practice, we introduce the derivative expansion to treat the non-locality of the potential,
\begin{equation}
  U({\bf r},{\bf r'}) = (V_0(r) + V_2(r) \nabla^2 + {\mathcal O}(\nabla^4)) \delta({\bf r-r'}),
\end{equation}
\fi
The effective leading order(LO) potential is given by
\begin{equation}
  V^{\rm LO}(r) = \frac{ \sum_{g\in O_h} R^{\dag}(g{\bf r},t) \left[ \dfrac{\nabla^2}{2 \mu} -\dfrac{\partial}{\partial t} + \dfrac{1}{8 \mu} \dfrac{\partial^2}{\partial t^2} \right] R(g{\bf r},t)}{\sum_{g\in O_h} R^{\dag}(g{\bf r},t) R(g{\bf r},t)},
\end{equation}
where invariance of the potential under the cubic rotation group $O_h$ is utilized to improve signals~\cite{Murano:2013xxa}.
In this study, we further determine the effective next-to-nexto-to-leading order (N$^2$LO) potential to extract resonance parameters more accurately.
The effective N$^2$LO potential $U^{\rm N^2LO}({\bf r},{\bf r'}) = \left( V_0^{\rm N^2LO} + V_2^{\rm N^2LO}  \nabla^2 \right) \delta({\bf r}-{\bf r'}) $ is determined by~\cite{Iritani:2018zbt}:
\begin{eqnarray}
  V_2^{\rm N^2LO}(r) &=& \frac{V_{\rho}^{\rm LO}(r) - V_{\pi \pi}^{\rm LO}(r)}{\nabla^2 R_{\rho}(r)/R_{\rho}(r) - \nabla^2 R_{\pi\pi}(r)/R_{\pi\pi}(r)} \label{eq:solutions_NLOlineqs_V2} \\
  V_0^{\rm N^2LO}(r) &=& V_{\rho}^{\rm LO}(r) - V_2^{\rm N^2LO}(r) \nabla^2 R_{\rho}(r)/R_{\rho}(r). \label{eq:solutions_NLOlineqs_V0}
\end{eqnarray}
where $R_i\ (i=\rho,\pi\pi)$ are the normalized correlation functions with $\rho$-type and $\pi\pi$-type source operators and $V^{\rm LO}_i(r)\ (i=\rho,\pi\pi)$ are the effective LO potentials obtained by $R_i\ (i=\rho,\pi\pi)$.

\section{Evaluation of correlation functions}
{
Since the P-wave $I=1$ $\pi \pi$ correlation functions contain quark creation$/$annihilation diagrams and momentum projection, we need all-to-all propagators to evaluate them.
Previous attempts of all-to-all calculation using the hybrid method reveal that it introduces large noise contamination originated from the stochastic estimations of the propagators.
Motivated by those lessons, we have explored an improved treatment of all-to-all propagators and found that the combination of the one-end trick, the sequential propagator technique, and the CAA can achieve both small noise contamination and small numerical cost.
}
Here we briefly explain one of the newly introduced techniques, the one-end trick, and outline how we combine them in evaluations of diagrams appearing in the correlation functions.

\subsection{The one-end trick}
Let us consider a combination of quark propagators given by
\begin{equation} \label{eq:oneend1}
  \sum_{\bf y} e^{i{\bf p \cdot y}} D^{-1}({\bf x}_1,t_1;{\bf y},t_0) \Gamma D^{-1}({\bf y},t_0;{\bf x}_2,t_2),
\end{equation}
where $D^{-1}$ is a quark propagator, $\Gamma$ is some product of gamma matrices, and $x_i = ({\bf x}_i,t_i)$ are arbitrary.
We abbreviate color and spin indices for simplicity.
The evaluation of such a combination naively needs two stochastic estimations for each, since each of them contains two all-to-all propagators.
The one-end trick, however, utilizes the $\gamma_5$-Hermiticity of the Dirac operator to estimate that structure with a single noise insertion as follows:
\begin{equation}
  \sum_{\bf y} D^{-1}({\bf x}_1,t_1;{\bf y},t_0) \Gamma D^{-1}({\bf y},t_0;{\bf x}_2,t_2) \approx \frac{1}{N_{\rm r}} \sum_{r=0}^{N_{\rm r}-1} \xi_{{\bf p},t_0[r]}({\bf x}_1,t_1) \otimes \chi_{\Gamma,t_0[r]}^{\dag} ({\bf x}_2,t_2) \gamma_5,
\end{equation}
with the ``one-end vectors''
\begin{eqnarray}
  \xi_{{\bf p},t_0[r]}(x) &\equiv& \sum_{\bf y} D^{-1}(x;{\bf y},t_0) \eta_{[r]}({\bf y}) e^{i{\bf p \cdot y}} \\
  \chi_{\Gamma,t_0[r]}(x) &\equiv& \sum_{\bf y} D^{-1}(x;{\bf y},t_0) \gamma_5 \Gamma^{\dag} \eta_{[r]}({\bf y}).
\end{eqnarray}
The one-end vectors $\xi$ and $\chi$ are obtained by solving the linear equation $D \xi = \eta e^{i{\bf p \cdot y}}$ and $D \chi = \gamma_5 \Gamma^{\dag} \eta$, respectively. The dilution technique\cite{Foley:2005ac} for noise reduction can be combined as well.
This trick is particularly suitable for the HAL QCD method since it does not introduce any stochastic estimations at the sink side, which otherwise strongly affects spatial dependences of the NBS wave function.
Moreover, a numerical cost and noise contamination are also reduced in accordance with a decrease in the number of noise vectors.

\subsection{Evaluation of diagrams}

\begin{figure}[htbp]
  \begin{tabular}{ccc}
  \begin{minipage}{0.3\hsize}
    \includegraphics[width=45mm,bb=0 0 287 272,clip]{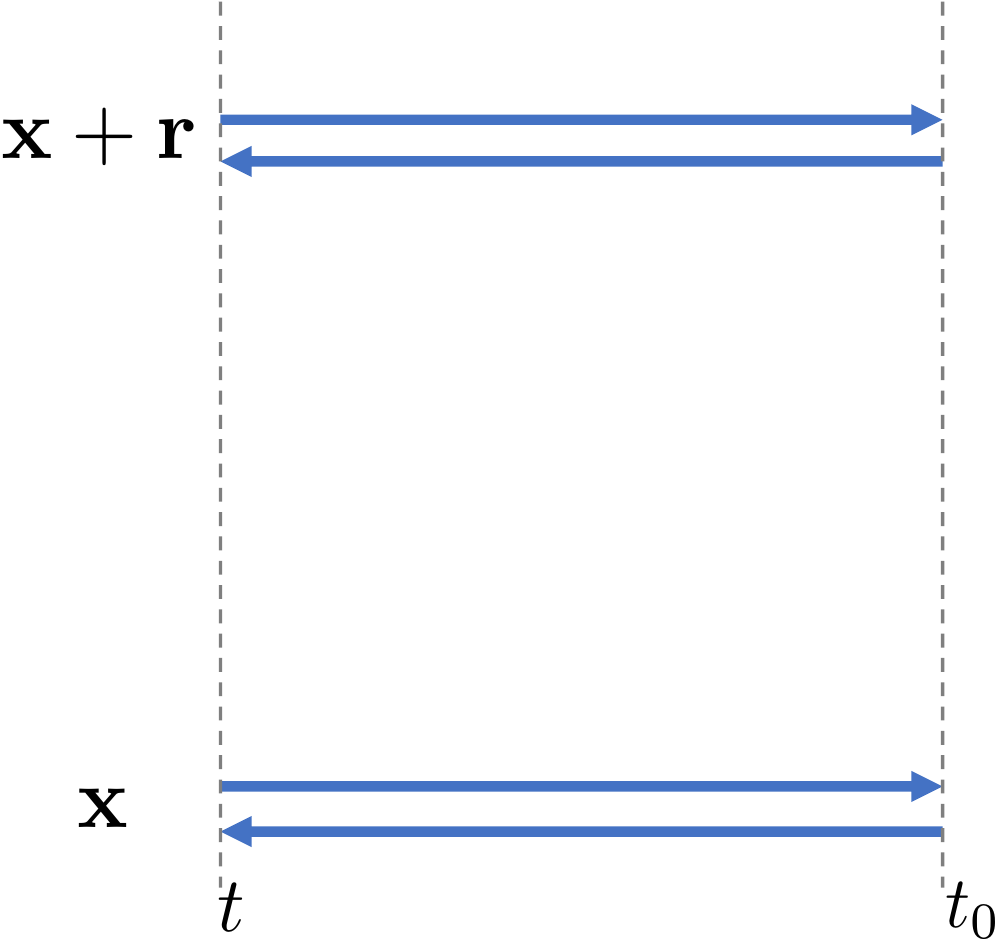}\\
    \centering
    separated diagram
  \end{minipage} &
  \begin{minipage}{0.3\hsize}
    \includegraphics[width=45mm,bb=0 0 293 271,clip]{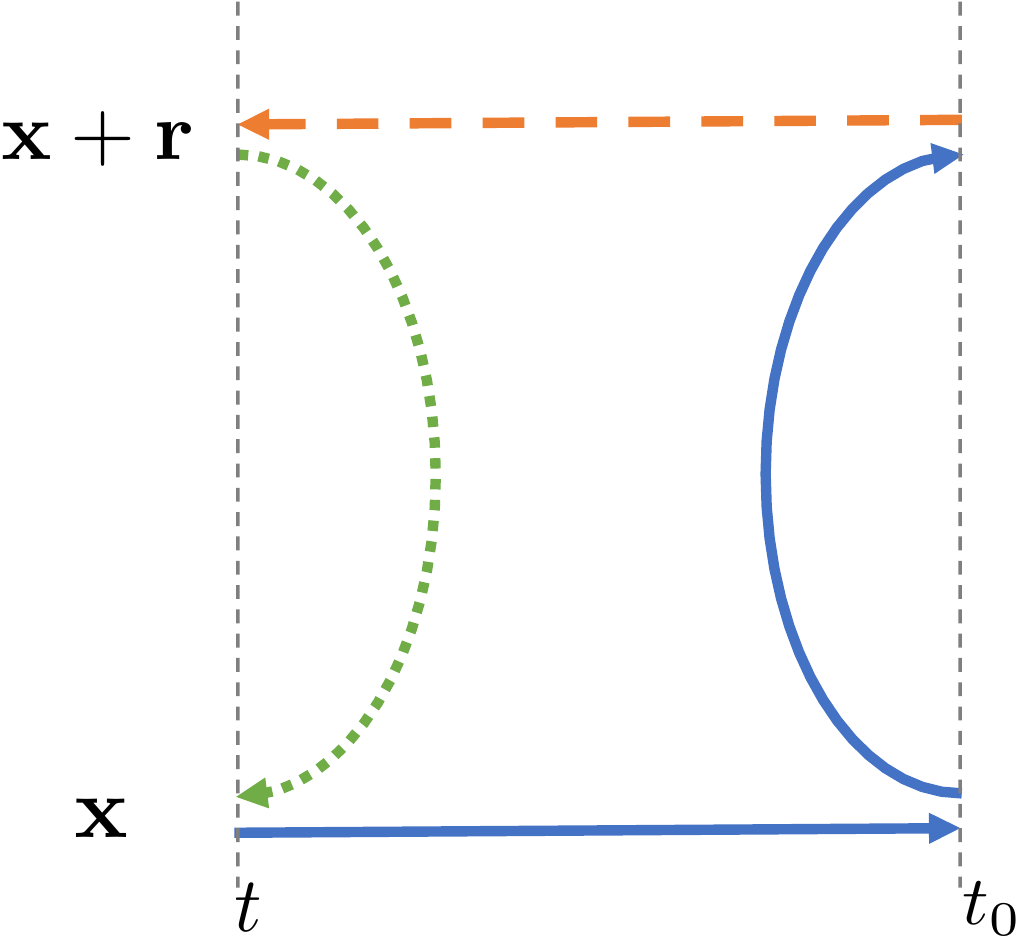}\\
    \centering
    box diagram
  \end{minipage} &
  \begin{minipage}{0.3\hsize}
    \includegraphics[width=45mm,bb=0 0 292 271,clip]{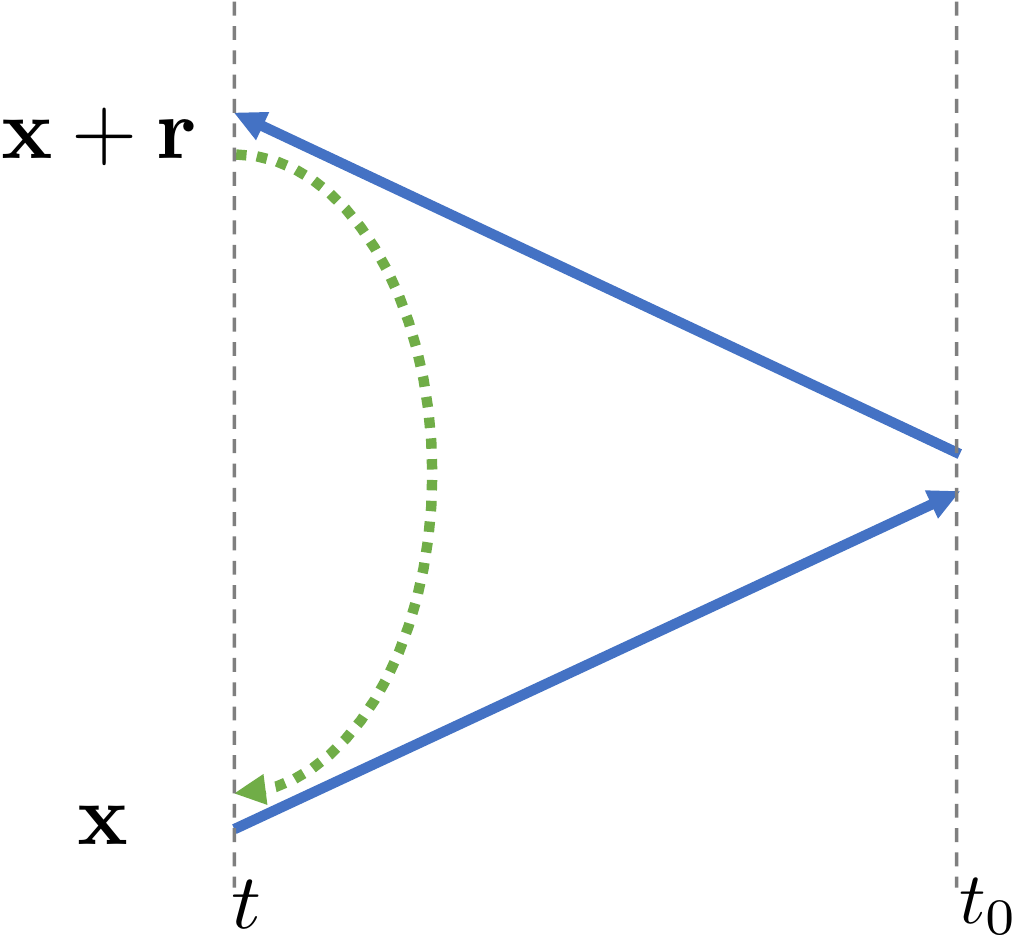}\\
    \centering
    triangle diagram
  \end{minipage}
\end{tabular}
\caption{Representative diagrams appear in correlation functions.
Blue solid, orange dashed and green dotted lines are quark propagators calculated by the one-end trick,
sequential propagator technique, and point-to-all propagator, respectively.
The CAA is applied to the coordinates ${\bf x}$ in the box and triangle diagram.
}
\label{fig:diagrams}
\end{figure}

%By combining the three techniques, we evaluate quark contraction diagrams with the least insertion of the stochastic estimators.
Figure~\ref{fig:diagrams} shows representative diagrams appearing in this study, where the techniques utilized in evaluations of quark propagators are shown by different colors and symbols.
The one-end trick and the sequential propagator calculation are utilized in the source part of the diagrams.
The CAA is applied to the center-of-mass coordinates between the two-pion operators at the sink part (${\bf x}$ in eq.~(\ref{eq:sinkop})), which has the translational invariance.
At the end of the day, we need at most two insertions of the stochastic estimators in our calculation and achieve about 10 times smaller statistical errors than previous attempts using the hybrid method.

\section{Numerical result}

We employ (2+1)-flavor full QCD configurations generated by the PACS-CS Collaborations~\cite{Aoki:2008sm} on a $32^3 \times 64$ lattice with the Iwasaki gauge action\cite{Iwasaki:1985we}
and a non-perturbatively improved Wilson-clover action\cite{Sheikholeslami:1985ij}.
In this setup, a lattice spacing is $a \approx 0.0907$ fm
and a pion mass $m_{\pi} \approx 0.41$ GeV, where the $\rho$ meson appears as a resonance with $m_{\rho} \approx 0.89$ GeV~\cite{Aoki:2011yj}.
To perform the exponential smearing at sink operators, we employ the Coulomb gauge fixing.
Table \ref{tab:setups_general} summarizes details of our simulations.
Furthermore, to remove the dominant higher partial wave contamination of $l=3$ originated from the reduced rotational symmetry, we apply the approximated partial wave decomposition recently introduced to lattice QCD\cite{Miyamoto:2019jjc}.
\if0
In lattice QCD, the rotational symmetry reduces to the cubic symmetry, and there exist higher partial wave components in the irreducible representation of the cubic group($l = 3,5,\cdots$ in this study).
This leads to systematic uncertainties in the HAL QCD potential, which exhibit as multi-valued structures of potentials as a function of $r$.
%To remove uncertainty coming from the dominant higher partial wave of $l=3$, we apply the approximated partial wave decomposition recently introduced to lattice QCD\cite{Miyamoto:2019jjc}.
We apply the approximated partial wave decomposition\cite{Miyamoto:2019jjc} to remove the dominant higher partial wave of $l=3$.
\fi
\begin{table}[tbp]
  \small
  \caption{Setups for statistics, the one-end trick and the CAA. $N_{\rm eig}$ and $N_{\rm ave}$ is the number of low eigenmodes and averaged points used in the CAA, respectively. The averaged points are chosen as ${\bf x} = (x_0 + 8l, y_0 + 8m, z_0 + 8n)\ {\rm mod}\ 32$ for $l,n,m \in \{0,1,2,3\}$, with a randomly chosen reference point ${\bf x}_0$ for each gauge configuration sample. Color and spinor dilutions are always used for noise vectors in the one-end trick.}
  \vspace{2mm}
  \centering
  \begin{tabular}{c|c|cc|cc}
    Source type & $N_{\rm conf} \times N_{\rm srctime}$ (Stat. error) & \multicolumn{2}{|c|}{One-end trick} & \multicolumn{2}{|c}{CAA} \\
    & & Noise vector & Space dilution & \ $N_{\rm eig}$ & $N_{\rm ave}$ \\ \hline \hline
    $\pi \pi$-type & $100\times64$ (jackknife with bin--size 5) & $Z_4$ noise & $s2$ (even-odd) & \ 300 & 64\\
    $\rho$-type & $200\times64$ (jackknife with bin--size 10) & $Z_4$ noise & $s4$ & \ 300 & 64\\
  \end{tabular}
  \label{tab:setups_general}
\end{table}

\subsection{LO and N$^2$LO potentials}
\begin{figure}[htbp]
  \hspace{-10mm}
  \begin{tabular}{cc}
  \begin{minipage}{0.5\hsize}
    \includegraphics[width=80mm,bb=0 0 828 581,clip]{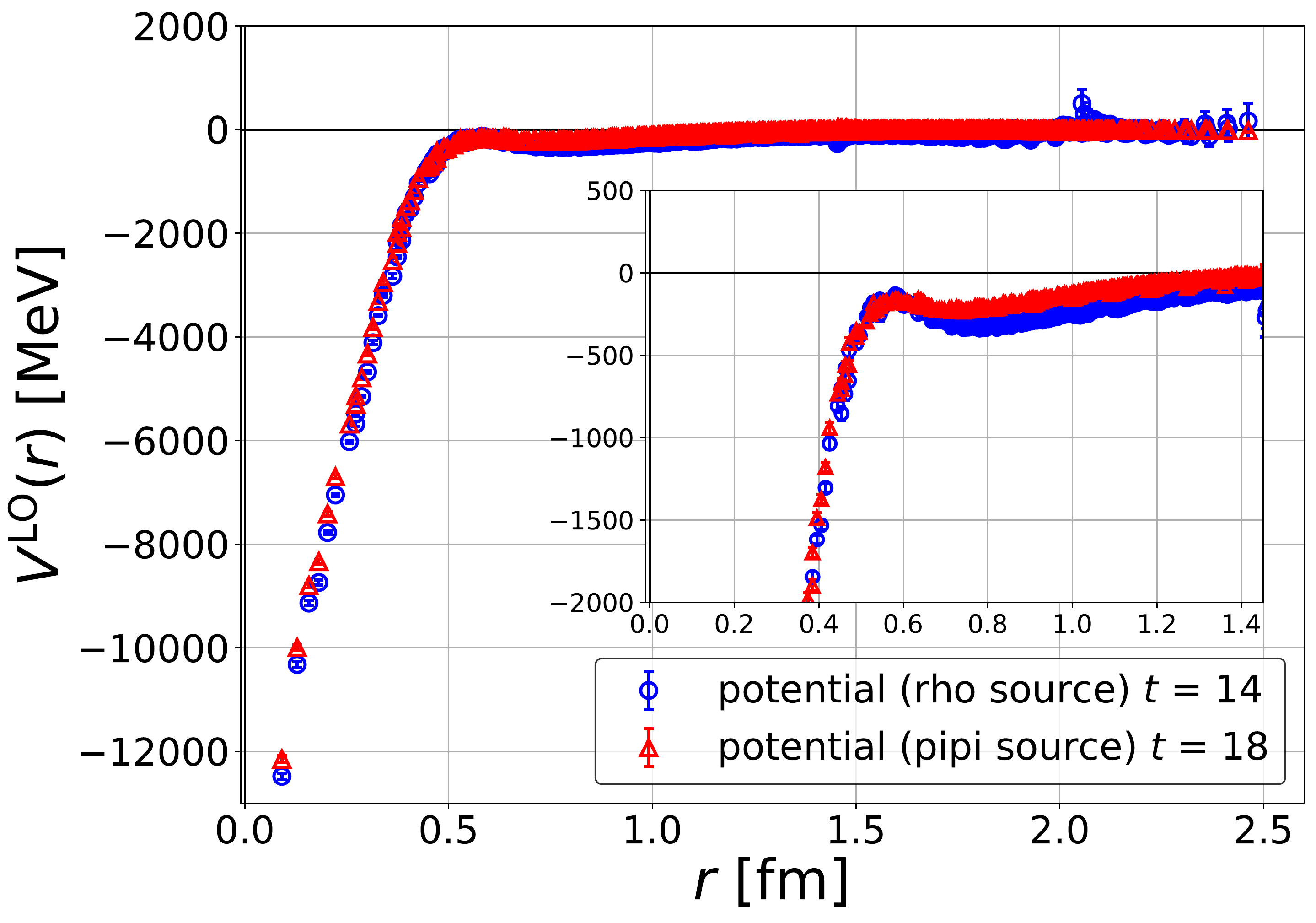}
  \end{minipage} &
  \begin{minipage}{0.5\hsize}
    \includegraphics[width=80mm,bb=0 0 815 581,clip]{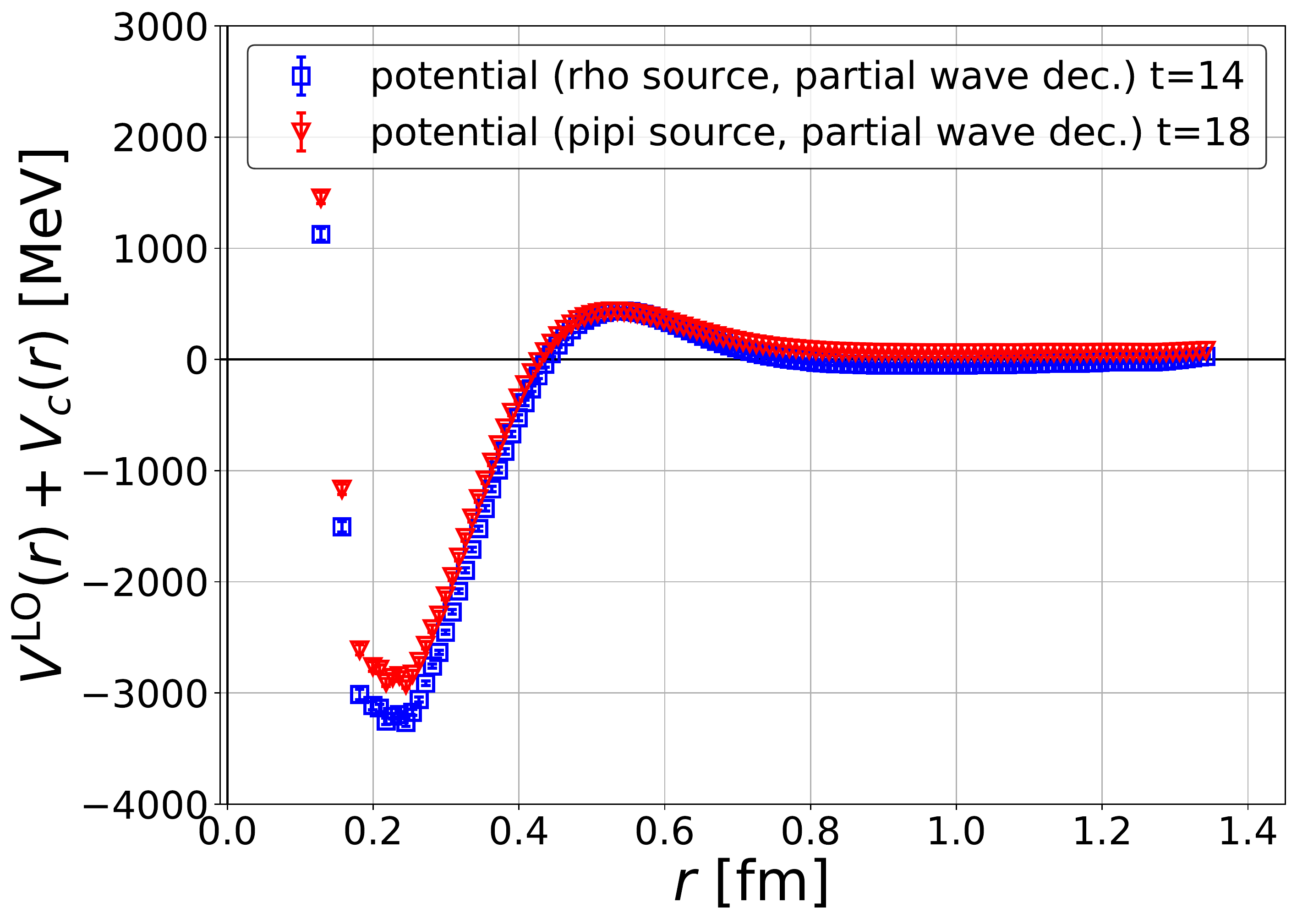}
  \end{minipage}
\end{tabular}
\caption{(Left) Effective LO potentials. Blue and red points show the results from the $\rho$-type source and the $\pi\pi$-type source, respectively. Inset shows an enlarged view of potentials. {(Right) Improved potentials obtained by the partial wave decomposition\cite{Miyamoto:2019jjc} with the P-wave centrifugal term, $V_c(r) = \frac{1}{2 \mu} \frac{1 \cdot 2}{r^2}$.}}
\label{fig:LOpotentials_t14}
\end{figure}
Figure~\ref{fig:LOpotentials_t14} (Left) shows the results for effective LO potentials.
We observed that the potentials are attractive at all distances.
We also notice that potentials obtained from different source operators are different from each other, suggesting a presence of non-negligible higher-order contributions in the derivative expansion.
Fig.~\ref{fig:LOpotentials_t14} (Right) gives
potentials after the partial wave decomposition with the P-wave centrifugal term added, which become much smoother as multi-valued structures are eliminated.
The potentials with the centrifugal term show characteristic features for the existence of a resonance state such as an attractive pocket at short distances and a potential barrier around $r=0.5$ fm.
\if0
To evaluate the physical observables, we fit the potentials with a multi-Gaussian shape,
\begin{equation} \label{eq:fitfunc_3G}
  V(r) = a_0 e^{-(r-a_1)^2/a_2^2} + a_3 e^{-(r-a_4)^2/a_5^2} + a_6 e^{-(r-a_7)^2/a_8^2}.
\end{equation}
The systematic uncertainty of physical observables are estimated by the uncertainty of the fits at small $r$.
\fi
To evaluate the physical observables, we fit the potentials with a multi-Gaussian shape, $V(r) = a_0 e^{-(r-a_1)^2/a_2^2} + a_3 e^{-(r-a_4)^2/a_5^2} + a_6 e^{-(r-a_7)^2/a_8^2}$.
The systematic uncertainty of physical observables is estimated by the uncertainty of the fits at small $r$.

\begin{figure}[tbp]
  \hspace{-10mm}
  \begin{tabular}{cc}
  \begin{minipage}{0.5\hsize}
    \includegraphics[width=80mm,bb=0 0 823 581,clip]{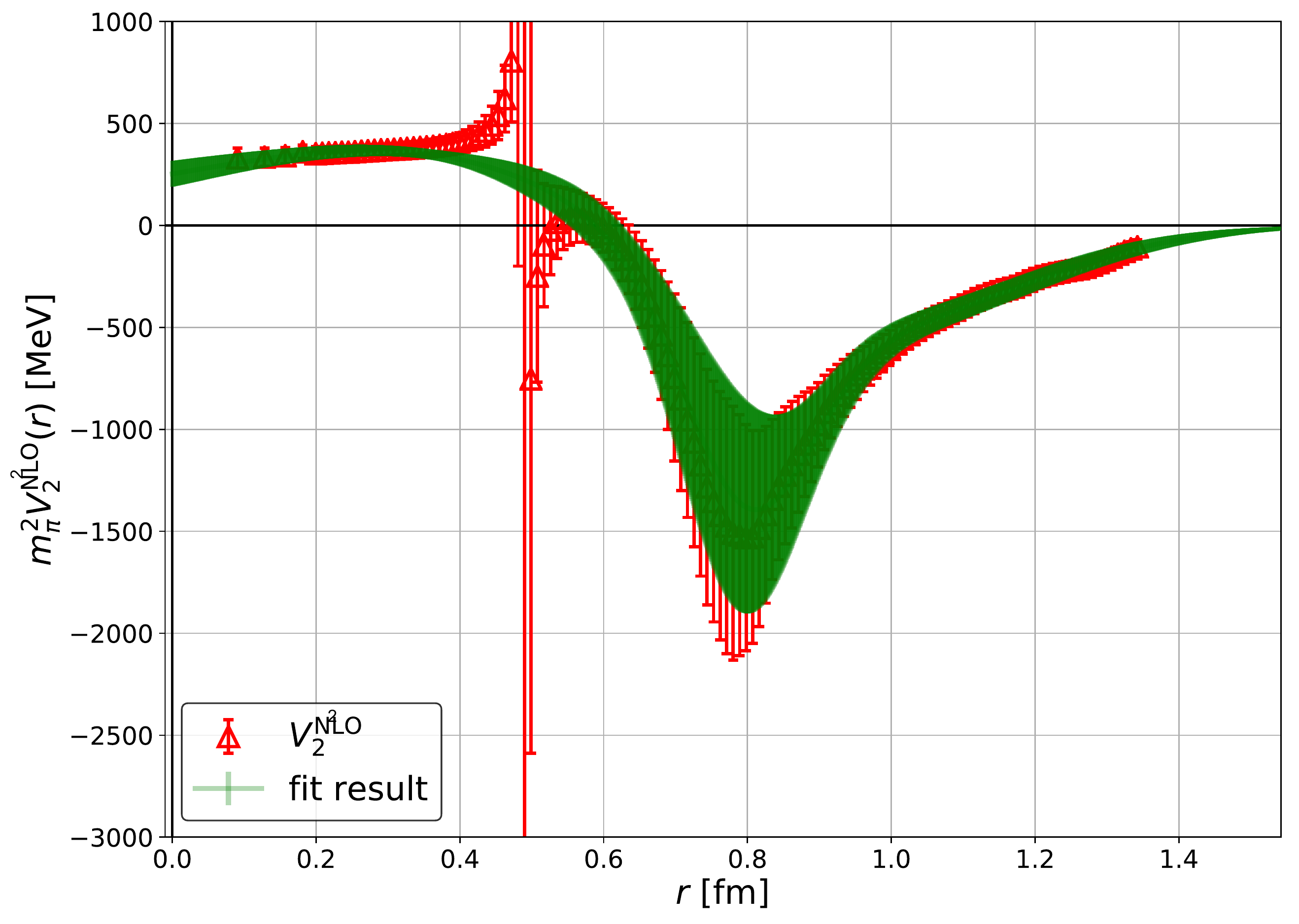}
  \end{minipage} &
  \begin{minipage}{0.5\hsize}
    \includegraphics[width=80mm,bb=0 0 822 581,clip]{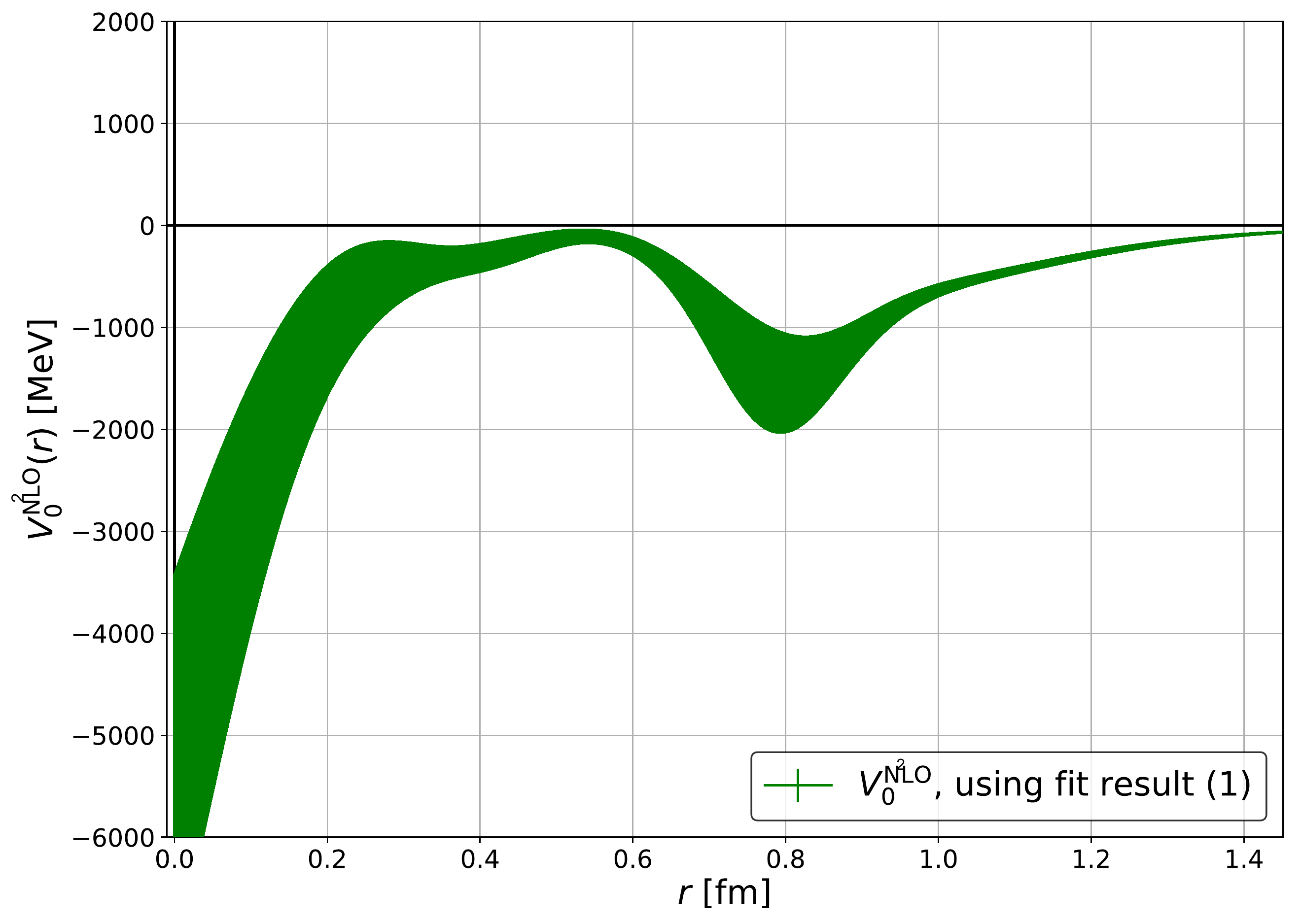}
  \end{minipage}
\end{tabular}
\caption{Effective N$^2$LO potentials. (Left) $V_2^{\rm N^2LO}$ obtained by the decomposed data (red triangles) and fit result (green band).
(Right) $V_0^{\rm N^2LO}$ obtained by fit results through the eq.(\ref{eq:solutions_NLOlineqs_V0}).
}
\label{fig:NLOpotentials}
\end{figure}

Next, let us consider the N$^2$LO potentials.
The red triangle plots in Fig.~\ref{fig:NLOpotentials} (Left) shows the $V_2^{\rm N^2LO}$ obtained by eq.(\ref{eq:solutions_NLOlineqs_V2}).
We observe a singular behavior at $r \approx 0.5$ fm, which comes from a vanishing denominator of $V_2^{\rm N^2LO}$.
{
In the fit of $V_2^{\rm N^2LO}$, we assume that the N$^4$LO or higher-order contribution in the derivative expansion can be neglected. This assumption leads us to employ a smooth fit function (same as the LO potentials) and use data points satisfying $1 - 2 \mu V_2^{\rm N^2LO} > 0$. The fit result is shown as a green band in Fig.~\ref{fig:NLOpotentials} (Left).
}
We then obtain $V_0^{\rm N^2LO}$ by combining all the fit results in eq.(\ref{eq:solutions_NLOlineqs_V0}), which are shown in Fig.\ref{fig:NLOpotentials} (Right).
Thanks to the new strategy to handle all-to-all propagators, we succeed in extracting the effective N$^2$LO potential of this system for the first time.

\subsection{Physical observables}
\begin{figure}[tbp]
  \hspace{-10mm}
  \begin{tabular}{cc}
  \begin{minipage}{0.5\hsize}
    \includegraphics[width=80mm,bb=0 0 806 583,clip]{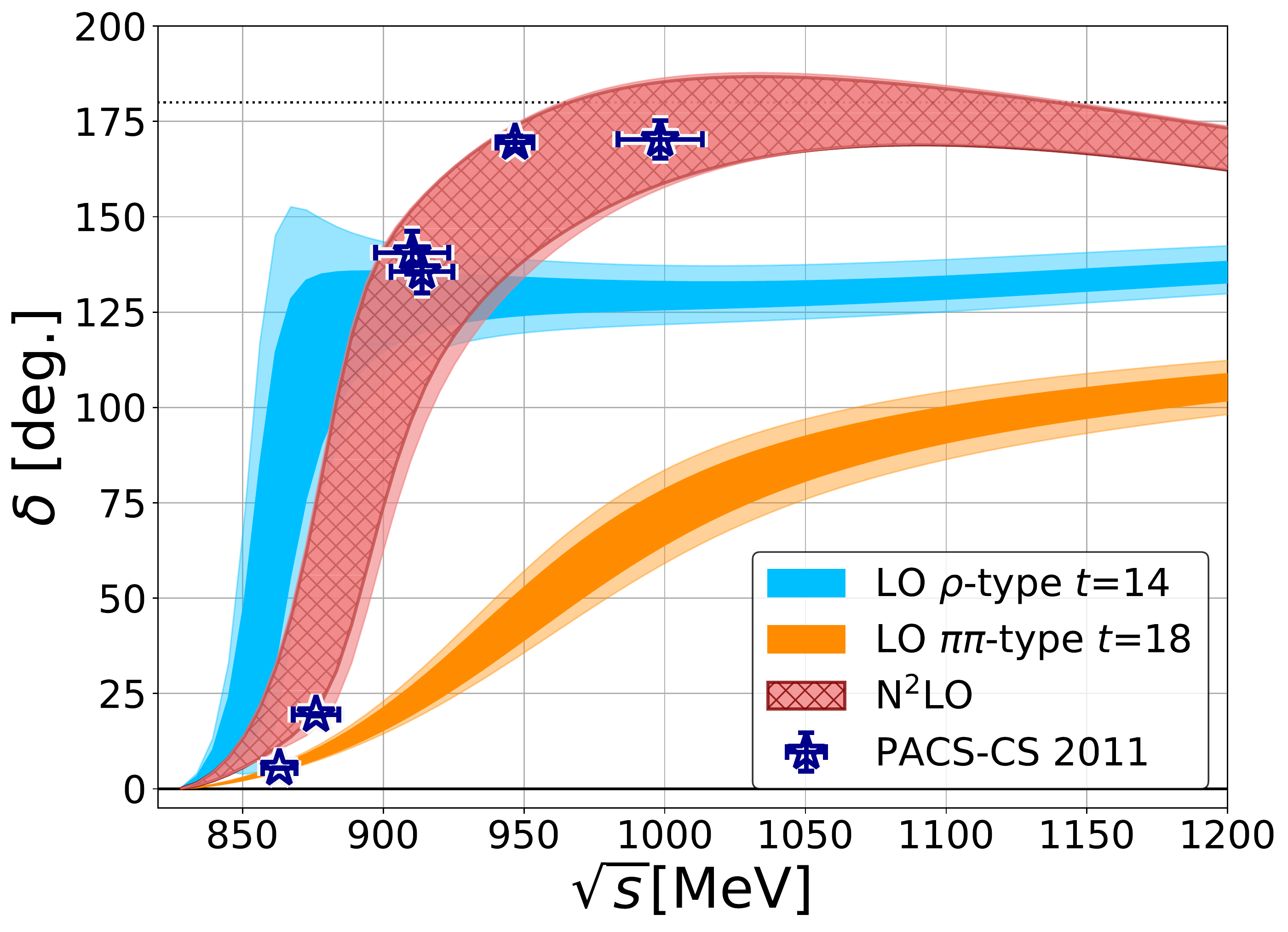}
  \end{minipage} &
  \begin{minipage}{0.5\hsize}
    \includegraphics[width=80mm,bb=0 0 759 576,clip]{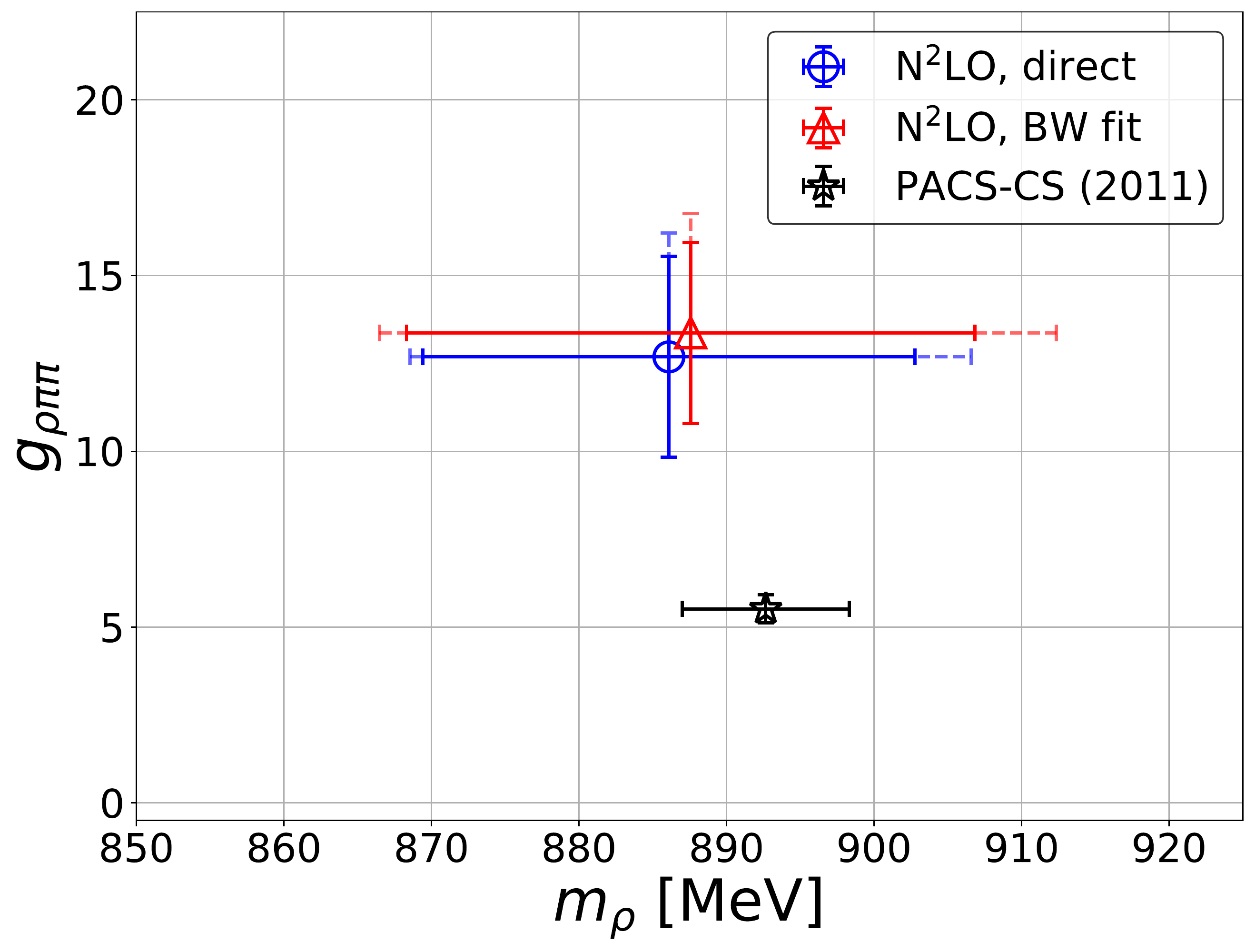}
  \end{minipage}
\end{tabular}
\caption{(Left) Scattering phase shift obtained by the LO and N$^2$LO potentials. Lighter color bands show the systematic uncertainty estimated by the uncertainty of the fits at small $r$. (Right) A comparison of N$^2$LO resonance parameters and previous result from PACS-CS(2011). Dotted bars give the systematic uncertainty. }
\label{fig:observables}
\end{figure}

Scattering phase shifts obtained by the LO and N$^2$LO potentials are summarized in Fig.\ref{fig:observables} (Left), together with the previous result from PACS-CS Collaboration\cite{Aoki:2011yj}.
All of our results show a typical resonant behavior, rising and crossing 90 degrees.
The LO results, however, largely deviate from the result of L\"uscher's method.
It indicates that the LO approximation is insufficient in our setup.
On the other hand, the N$^2$LO result becomes roughly consistent with the PACS-CS result.
The remaining difference between the N$^2$LO and the PACS-CS result observed in the low-energy region can be understood as follows:
Our calculations are performed only in the center-of-mass frame,
where the corresponding energy levels on the current lattice volume do not cover the low-energy region near the $\pi\pi$ threshold.
Therefore, the N$^2$LO approximation could suffer from the large truncation error of the derivative expansion in such a low-energy region.

Finally, we extract parameters of the $\rho$ resonance using the N$^2$LO potential.
We employ two different procedures, fit of the Breit-Wigner parametrization and direct pole search of the S-matrix.
Extracted parameters are summarized in Fig.\ref{fig:observables} (Right).
The resonance mass $m_{\rho}$ is consistent with the previous study, but the coupling $g_{\rho \pi \pi}$ is twice as large.
We suspect that this difference mainly comes from the discrepancy in a low-energy region as mentioned above.
While a resonance mass is likely to be well reproduced as long as the resonance appears
in the energy region accessible in the center-of-mass frame,
the coupling may suffer from larger systematics
since it is sensitive to energy dependence on a much wider range around the resonance.
This observation gives us a useful lesson
for the study of P-wave (or higher partial wave) resonances by the HAL QCD method
with the center-of-mass frame.
If the non-locality of the potential happens to be large,
the truncation errors could be large at low-energies near the threshold.
To control the systematics appearing in the determination of the effective coupling and decay width,
one may, for example, introduce the laboratory frame calculations in the analysis
or tune lattice parameters (box size etc.) to cover an energy region of a target resonance only by the center-of-mass frame spectra.

\section{Summary}

We study the P-wave $I=1$ $\pi\pi$ interaction at $m_{\pi} \approx 0.41$ GeV, where the $\rho$ meson appears as a resonance with $m_{\rho} \approx 0.89$ GeV.
%To achieve a high-precision evaluation of quark contraction diagrams including all-to-all propagators,
We newly introduce a combination of the one-end trick, the sequential propagator technique, and the CAA in the evaluation of correlation functions.
Thanks to those techniques, we successfully determine the HAL QCD potential at the N$^2$LO of the derivative expansion for the first time and find the pole structure corresponding to the $\rho$ resonance.
The resonance mass is consistent with the previous result by L\"uscher's method, but somewhat larger coupling and decay width are obtained.
We suspect that this discrepancy comes from the large systematics in the low-energy region far from the center-of-mass energy spectra, which may be reduced by, for instance, introducing the laboratory frame calculations.
Although there remain some issues to be investigated, this study shows that we can study hadronic resonances requiring all-to-all calculations at the N$^2$LO level in the HAL QCD method.

\acknowledgments
The authors thank members of the HAL QCD Collaboration for fruitful discussions.
We thank the PACS-CS Collaboration~\cite{Aoki:2008sm} and ILDG/JLDG~\cite{Amagasa:2015zwb} for providing their configurations.
The numerical simulation in this study is performed on the HOKUSAI Big-Waterfall in RIKEN and the Oakforest-PACS in Joint Center for Advanced HighPerformance Computing (JCAHPC).
The framework of our numerical code is based on Bridge++ code~\cite{Ueda:2014rya} and its optimized version for the Oakforest-PACS by Dr. I. Kanamori~\cite{Kanamori:2018hwh}.
This work is supported in part
by HPCI System Research Project (hp200108, hp210061),
by the Grant-in-Aid of the Japanese Ministry of Education, Sciences and Technology, Sports and Culture (MEXT) for Scientific Research (Nos.~JP16H03978,  JP18K03620, JP18H05236, JP18H05407, JP19K03847)
and by the Grant-in-Aid for the Japan Society for the Promotion of Science (JSPS) Fellows (No. JP20J11502).
Y.A. is supported in part by JSPS.
S. A. and T. D. are also supported in part
by a priority issue (Elucidation of the fundamental laws and evolution of the universe) to be tackled by using Post ``K" Computer,
by Program for Promoting Researches on the Supercomputer Fugaku (Simulation for basic science: from fundamental laws of particles to creation of nuclei),
and by Joint Institute for Computational Fundamental Science (JICFuS).

\if0
\bibliography{ref}
\bibliographystyle{JHEP.bst}
\fi

\end{document}